\documentclass[useAMS,usenatbib]{mn2e}

\usepackage{booktabs,multicol}
\usepackage{color}
\usepackage{multirow}
\usepackage{braket}
\usepackage{graphicx}
\usepackage{cite}
\usepackage{bm}

\title[Photoionization of Cl-like Argon]{Valence and L-shell photoionization of Cl-like Argon using R-matrix techniques}
\author[N. B. Tyndall, C. A. Ramsbottom, C. P. Ballance and A. Hibbert]{N. B. Tyndall$^{1}$\thanks{E-mail:
ntyndall01@qub.ac.uk}, C. A. Ramsbottom$^{1}$, C. P. Ballance$^{1}$ and A. Hibbert$^{1}$\\
$^{1}$School of Mathematics \& Physics,\\
The Queen's University of Belfast, Belfast,\\
BT7 1NN,\\
Northern Ireland\\}
\begin{document}

\date{Accepted (date). Received (date); in original form (date)}

\pagerange{\pageref{firstpage}--\pageref{lastpage}} \pubyear{2002}

\maketitle

\label{firstpage}


\begin{abstract}
Photoionization cross sections are obtained using the relativistic Dirac Atomic \textbf{R}-matrix Codes ({\sc darc}) for all valence and L-shell energy ranges between 27-270eV. A total of 557 levels arising from the dominant configurations 3s$^2$3p$^4$, 3s3p$^5$, 3p$^6$, 3s$^2$3p$^3$[3d, 4s, 4p], 3p$^5$3d, 3s$^2$3p$^2$3d$^2$, 3s3p$^4$3d, 3s3p$^3$3d$^2$ and 2s$^2$2p$^5$3s$^2$3p$^5$ have been included in the target wavefunction representation of the Ar {\sc iii} ion, including up to 4p in the orbital basis. We also performed a smaller Breit-Pauli ({\sc bp}) calculation containing the lowest 124 levels. Direct comparisons are made with previous theoretical and experimental work for both valence shell and L-shell photoionization. Excellent agreement was found for transitions involving the $^2$P$^{\rm o}$ initial state to all allowed final states for both calculations across a range of photon energies. A number of resonant states have been identified to help analyze and explain the nature of the spectra at photon energies between 250 and 270eV.
\end{abstract}




\section{Introduction}\label{sec:introduction}
The complex process of photoionization in such fields as plasma physics and astrophysics is of ongoing interest. Photo-processes are clearly evident in astrophysics for spectral modelling. However, hydrogenic approximations are often employed where data is absent or incorrectly formatted \citep{2012A&A...546A..28J}. The Opacity Project (OP) \citep{1992RMxAA..23..107C, 1993A&A...275L...5C} provides one source for photoionization cross sections which are often beneficial, but generally more sophisticated calculations need to be performed. A relativistic approach can offer an accurate level resolved spectra, allowing detailed comparisons to be carried out between experiment and theory \citep{2002JPhB...35L.137M, 2009JPhB...42w5602M}.

Experiments have recently and are currently being performed for inner shell transitions. Examples of the excellent agreement achieved between theory and experiment has been noted for the lighter systems B$^{2+}$ and N$^{+}$ \citep{2010JPhB...43m5602M, 2011JPhB...44q5208G} during K-shell photoionization. In addition, poor agreement at higher photon energies was found between OP results and experimental photoionization cross sections for C$^{+}$ \citep{2001ApJS..135..285K}. This reinforces the need to conduct sophisticated \textbf{R}-matrix calculations on high performance computing facilities to match advancements in experimental technologies.

We focus in this paper on the photoionization of Ar {\sc ii}. The strong $\lambda 7135 {\rm \AA}$ line (from the neighbouring ion stage Ar {\sc iii}), which is a consequence of the forbidden transition $^1$D$_2$ $\rightarrow$ $^3$P$_2$ within the ground state complex 3s$^2$3p$^4$, has been measured in a number of planetary nebula \citep{1987ApJS...65..405A}. It has also been suggested that Argon be used as a metal abundance indicator for spectral diagnostics \citep{1980ApJ...240...99B}. This intense line has been observed in the binary of $\eta$ Carinae and used to model wind interaction within the system \citep{2009MNRAS.396.1308G}. Specifically, there has been work adopting the line ratio (Ar {\sc iii} $\lambda 7135 {\rm \AA}$)/(O {\sc iii} $\lambda 5007 {\rm \AA}$) to determine Oxygen abundance in H {\sc ii} and star forming regions \citep{2006A&A...454L.127S}. Another useful application is in the analysis of the ionic/elemental abundances of Ar$^{2+}$/Ar in three H {\sc ii} regions using data recorded from \textit{Spitzer} \citep{2008ApJ...680..398L}.

A number of lines have also been detected using beam-foil techniques in the wavelength region $\lambda$500-$\lambda$1000$ {\rm \AA}$ \citep{PINNINGTON:71}, theta pinch experiments \citep{1987JPhB...20..693H} for the optical region, and also through capillary-discharge tube experiments \citep{2000BrJPh..30..386L}. A compilation of weighted oscillator strengths have been evaluated through a Multi-Configuration Hartree-Fock {\sc mchf} approach in two separate cases \citep{2001JQSRT..69..171L, 2006ADNDT..92..607F}. Many of these transitions are only accessible by including the 3d and $n=4$ complex orbitals and retention of these are critical for subsequent calculations.

Lowly ionized Argon systems, and even neutral argon have been studied with scrutiny due to their importance as mentioned above. We focus on the work that has been carried out by \citet{2011PhRvA..84a3413C} and \citet{2012PhRvA..85d3408B}, which consider the photoionization process involving Ar$^+$. Both works include experimental and theoretical calculations. 

\citet{2011PhRvA..84a3413C} have computed absolute cross sections for the valence shell photoionization up to photon energies of 60eV for the statistically weighted, ground state complex only. All experimental results are obtained from the merged beam technique that has been carried out at the Advanced Light Source (ALS) with a spectral resolution of 10meV. In contrast, \citet{2012PhRvA..85d3408B} have investigated photoionization at the L-shell between an energy range of 240-282eV for multiple ion stages of Argon. The experiment has been conducted at SOLEIL in France to a larger spectral resolution of 140meV and directly compares with {\sc mchf} and {\sc opas} calculations detailed therein. The experimental procedure is similar to that mentioned above, with the photon beam produced by a magnet undulator on the PLEIADES line, merged with various argon ion beams. 

This paper details single photon valence shell and L-shell photoionization processes, comparing with previous work where appropriate. The remainder of this paper shall be structured as follows. Section \ref{sec:structure} discusses the structure calculation in preparing a model for Ar {\sc iii}, Section \ref{sec:rmatrix} contains a brief overview on \textbf{R}-matrix theory applied to photoionization and Section \ref{sec:results} presents the results obtained and discussions. Finally in Section \ref{sec:conclusions} we summarize our findings in the conclusions.



\section{Structure model}\label{sec:structure}
The photoionization processes of interest can be described by the following equations,
\begin{equation}\label{eq:ground}
h\nu + (2p^63s^23p^5) ^2\rm{P}^{o}_{3/2, 1/2} \rightarrow Ar^{2+} + \rm{e}^-
\end{equation}
\begin{equation}\label{eq:excited}
h\nu + (2p^63s3p^6) ^2\rm{S}^{e}_{1/2} \rightarrow Ar^{2+} + \rm{e}^-
\end{equation}
where it is found that the dominant contributions to the total photoionization come from the Ar$^{2+}$ 3s$^2$3p$^4$ and 3s3p$^5$ levels. We have investigated two methods for generating an appropriate basis set expansion of the Ar {\sc iii} ion. The first is carried out through a Breit-Pauli approach using the computer code {\sc civ3} \citep{1975CoPhC...9..141H, 1991CoPhC..64..455H}, and secondly, using the relativistic computer code {\sc grasp0} \citep{1996CoPhC..94..249P}. This stage of the calculation is crucially important enabling an accurate representation of both the initial target as well as the residual ion which is then to be constructed and incorporated into the \textbf{R}-matrix method. 

\begin{table}
\begin{center}
\begin{tabular}{@{} l *3c @{}}
\toprule
\multicolumn{1}{c}{Calculation} & Number of & Configurations \\
 & levels & included \\
 \midrule
       \multicolumn{1}{c}{PBP} & 124 &  3s$^2$3p$^4$, 3s3p$^5$, 3p$^6$, 3s$^2$3p$^3$3d\\
       \multicolumn{1}{c}{} & & 3s$^2$3p$^3$[4s, 4p, 4d] \\  
       
        \multicolumn{1}{c}{} & & \\
                     
         \multicolumn{1}{c}{DARC1} & 209 & 3s$^2$3p$^4$, 3s3p$^5$, 3p$^6$, 3s$^2$3p$^3$3d\\
           \multicolumn{1}{c}{} &  & 3s$^2$3p$^3$[4s, 4p] +  \\
           \multicolumn{1}{c}{} &  & 3s$^2$3p$^2$3d$^2$ + 3p$^5$3d \\
                                          
             \multicolumn{1}{c}{} & & \\                     
                                                        
            \multicolumn{1}{c}{DARC2} & 257 &  DARC1 + \\
             \multicolumn{1}{c}{} & &  3s$^2$3p$^3$[4d, 5s] \\

               \multicolumn{1}{c}{} & & \\

             \multicolumn{1}{c}{DARC3} & 557 &  DARC1 + \\
             \multicolumn{1}{c}{} & &  3s3p$^4$3d + 3s3p$^3$3d$^2$ \\
                    \multicolumn{1}{c}{} & & + 2s$^2$2p$^5$3s$^2$3p$^5$ \\

      \bottomrule
 \end{tabular}
 \caption{The list of calculations performed throughout this paper are recorded and indexed for reference in the first column. The configurations and levels associated are also retained. \label{tab:calculations}}
 \end{center}
\end{table}

\subsection{Breit-Pauli approach}
We employed an analytic Slater type orbital description for the bound orbitals up to 3p from the tables of \citet{1974ADNDT..14..177C}. The computer package {\sc civ3} was then utilized to extend this basis expansion by including the 3d, 4s, 4p and 4d orbitals. These additional orbitals have been optimised in an $LS\pi$ coupling scheme on the lowest quintet states of the configurations 3s$^2$3p$^3$[3d, 4s, 4p, 4d] respectively. A total of 124 $J\pi$ levels were included in the basis set with configurations from 3s$^2$3p$^4$, 3s3p$^5$, 3p$^6$ and 3s$^2$3p$^3$[3d, 4s, 4p, 4d]. Configuration-interaction terms are also included to account for additional correlation in each wavefunction. These configuration-interaction expansions of the target wavefunctions employ a Breit-Pauli approach through one body perturbative corrections to the non-relativistic Hamiltonian operator. These corrections are described in full in the literature \citep{1980JPhB...13.4299S} and carried through to be used consistently in the Breit-Pauli (PBP) \textbf{R}-matrix method. This, the first of our Ar {\sc iii} models, is labelled in Table \ref{tab:calculations} as PBP.

\subsection{Relativistic approach}
The computer code {\sc grasp0} has also been used to construct a bound orbital basis set for Ar {\sc iii}. The method involves the Dirac-Coloumb Hamiltonian,
\[
H_{D}= \sum_i -ic\bm{\alpha} \nabla_i + (\bm{\beta} - 1)c^2-\frac{Z}{r_i} + \sum_{i<j}\frac{1}{|\mathbf{r}_j-\mathbf{r}_i|}
\]
where the electrons are labelled by $i$ and $j$ and the summation is taken over all electrons of the system. The matrices $\bm{\alpha}$ and $\bm{\beta}$ are directly related to the Pauli spin matrices, $c$ is the speed of light and the atomic number is $Z=18$. The relativistic orbitals are described with a large component, $\mathcal{P}_{nl}$ and small component $\mathcal{Q}_{nl}$. The target wavefunctions are appropriately defined on a radial grid for input into the relativistic Dirac Atomic \textbf{R}-matrix Codes ({\sc darc}).

Unlike in {\sc civ3} where we optimise the additional orbitals on the lowest lying states of the respective configuration, {\sc grasp0} considers the optimization process on every state included in the calculation, unless specified otherwise by the user.

\begin{table*}
\begin{center}
\begin{tabular}{@{} l *9r *1c @{}}
\toprule
\multicolumn{1}{c}{Index} & Configuration & Level & NIST & \multicolumn{4}{c}{\textbf{Present}} & Stancalie & Burgos  \\
 & & & & PBP & DARC1 & DARC2 & DARC3 & et al. & et al.  \\
 \midrule
 
       \multicolumn{1}{c}{1} & 3s$^2$3p$^4$ & $^3$P$_2$ & 0.0000 & 0.0000 & 0.0000 & 0.0000 & 0.0000 & 0.0000 & 0.0000 \\
        \multicolumn{1}{c}{2} & 3s$^2$3p$^4$ &$^3$P$_1$ & 0.1379 & 0.1498 & 0.1391 & 0.1391 &  0.1367 & 0.1325 & 0.1306\\
        \multicolumn{1}{c}{3} & 3s$^2$3p$^4$ &$^3$P$_0$ & 0.1947 & 0.2119 & 0.1952 & 0.1952 & 0.1942 & 0.1895 & 0.1864\\
       \multicolumn{1}{c}{4} & 3s$^2$3p$^4$ & $^1$D$_2$ & 1.7370 & 2.0093 & 2.1402 & 2.1402 & 2.0423 & 1.9266 & 2.0245\\
       \multicolumn{1}{c}{5} & 3s$^2$3p$^4$ & $^1$S$_0$ & 4.1244 & 4.3337 & 3.4318 & 3.4318 & 4.2673 & 4.3089 & 3.8654\\ 
      \multicolumn{1}{c}{6} & 3s3p$^5$ & $^3$P$_2$ & 14.1095 & 14.1116 & 14.1877 & 14.0961 & 13.9520 & 13.9529 & 13.6370\\ 
        \multicolumn{1}{c}{7} & 3s3p$^5$ & $^3$P$_1$ & 14.2331 & 14.2461 & 14.3136 & 14.2219 & 14.0752 & 14.2220 & 13.7526\\ 
         \multicolumn{1}{c}{8} & 3s3p$^5$ & $^3$P$_0$ & 14.2988 & 14.3164 & 14.3793 & 14.2877 & 14.1394 & 14.1490 & 13.8125\\ 
        \multicolumn{1}{c}{9} & 3s3p$^5$ & $^1$P$_1$ & 17.8565 & 18.4640 & 18.2749 & 18.2060 & 18.1696 & 17.4928 & 17.7255\\
          \multicolumn{1}{c}{10} & 3s$^2$3p$^3$3d & $^5$D$_0$ & -- & 18.2182 & 18.0206 & 18.0102 &  17.6992 & -- & --\\   
         \multicolumn{1}{c}{11} & 3s$^2$3p$^3$3d & $^5$D$_1$ & 17.9635 & 18.2203  & 18.0211 & 18.0107 & 17.6996 & 17.5912 & 17.9119\\   
          \multicolumn{1}{c}{12} & 3s$^2$3p$^3$3d & $^5$D$_2$ & 17.9642 & 18.2243 & 18.0220 & 18.0116 & 17.7005 & 17.5908 & -- \\   
         \multicolumn{1}{c}{13} & 3s$^2$3p$^3$3d & $^5$D$_3$ & 17.9650 & 18.2306 & 18.0233 & 18.0130 & 17.7019 & 17.3458 & 17.9214 \\   
          \multicolumn{1}{c}{14} & 3s$^2$3p$^3$3d & $^5$D$_4$ & 17.9667 & 18.2394 & 18.0255 & 18.0152 & 17.7040 & 17.5911 & 17.9296\\        
\bottomrule
 \end{tabular}
 \caption{Energies and assignments for the lowest 14 levels of the Ar {\sc iii} system presented in eV. \citet{2012EPJD...66...84S} and  \citet{2009A&A...500.1253M} are theoretical \textbf{R}-matrix results for electron impact excitation calculations, and the NIST results are observational data taken from \citet{2010JPCRD..39c3101S}. The remaining results are the present works summarized in Table \ref{tab:calculations}. \label{tab:energy}}
 \end{center}
\end{table*}

Initially we have included the important configurations 3s$^2$3p$^4$, 3s3p$^5$, 3p$^6$, 3s$^2$3p$^3$[3d, 4s, 4p], 3p$^5$3d and 3s$^2$3p$^2$3d$^2$, which gives rise to 209 levels. This model is labelled as DARC1 in Table \ref{tab:calculations}. We augment this model with the inclusion of the 3s$^2$3p$^3$[4d, 5s] configurations in DARC2 raising the number of levels to 257. The reason to perform this slightly larger evaluation was to test whether the inclusion of these high lying $nl=$ 4d and 5s levels affect the convergence of the photoionization cross section, or whether their contribution could be deemed negligible. An accurate representation for the low-lying wavefunctions of the residual ion is always of major importance for photoionization calculations. In an attempt to further improve correlation effects we perform a final relativistic evaluation, in which we include the additional 3s3p$^4$3d levels (mixing with 3s$^2$3p$^4$ and have the effect of lowering the relative $^3$P$_2$ ground energy), as well as all levels with configuration 3s3p$^3$3d$^2$ which improve the odd parity levels. The configuration 2p$^5$3s$^2$3p$^5$ has also been incorporated into the expansion of Ar {\sc iii} as it allows us to extend our evaluations to L-shell photoionization and results in an additional 10 levels. We label this, our largest Ar {\sc iii} model, as DARC3 in Table \ref{tab:calculations} containing 557 individual fine-structure levels. 


\begin{figure*}
\includegraphics[scale=0.6, angle=-90]{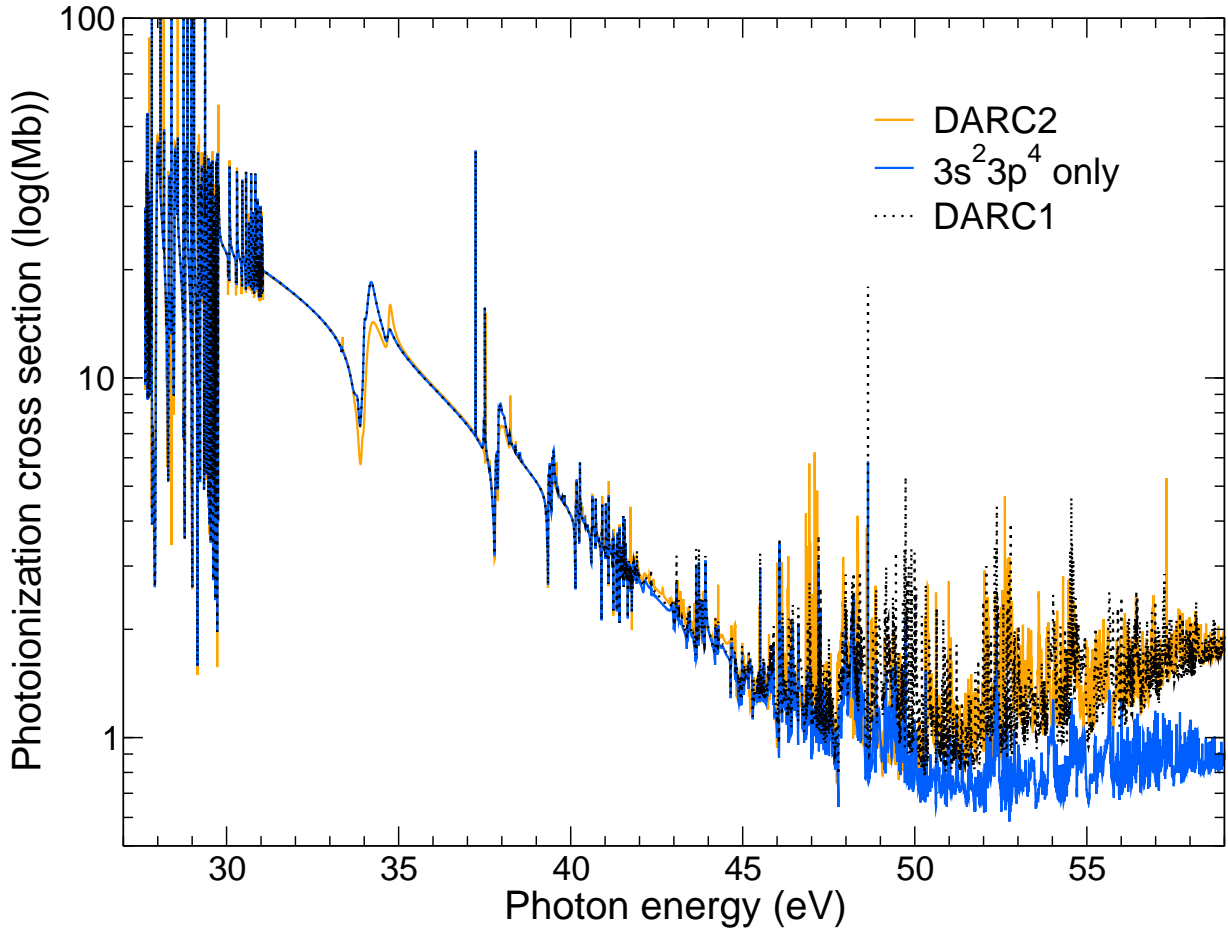}
\caption{Photoionization cross section from the initial $^2$P$^{\rm{o}}_{3/2}$ to allowed final states given in Mb on a logarithmic scale against the photon energy in eV. The dashed black line represents the result from DARC1, the solid blue line is the contribution from levels indexed 1-5 in Table \ref{tab:energy}, and the solid orange line is the extension to DARC2. \label{fig:comparison}}
\end{figure*}

We present in Table \ref{tab:energy} the energy levels (in eV) relative to the Ar {\sc iii} ground state for the lowest 14 fine-structure levels for this ion. We directly compare the {\em ab initio} energies from the PBP, DARC1, DARC2 and DARC3 evaluations with two theoretical \textbf{R}-matrix works \citep{2012EPJD...66...84S, 2009A&A...500.1253M}, both of which performed electron-impact excitation evaluations and generated their basis set for Ar {\sc iii} with the {\sc autostructure} \citep{1986JPhB...19.3827B}, adapted from the original {\sc superstructure} \citep{1974CoPhC...8..270E}, computer package. Comparisons are also made with the recorded NIST levels compiled by \citet{2010JPCRD..39c3101S} which incorporates designations from observed spectral analysis of existing works. 

All of the present calculations agree extremely well when compared with NIST, but the results from the DARC3 model give best overall agreement across the 14 levels considered. Differences of less than 4\% are found for all energy separations with the exception of the 3s$^2$3p$^4\; ^1$D$_2$ state where a difference of 15\% is recorded. This level of disparity is evident, however, for all the theoretical predictions listed. Due to the sophistication of the DARC3 model and the fact that it allows us to investigate L-shell photoionization, it is this model that we incorporate primarily into our collision calculations.

\begin{figure*}
\includegraphics[scale=0.6, angle=-90]{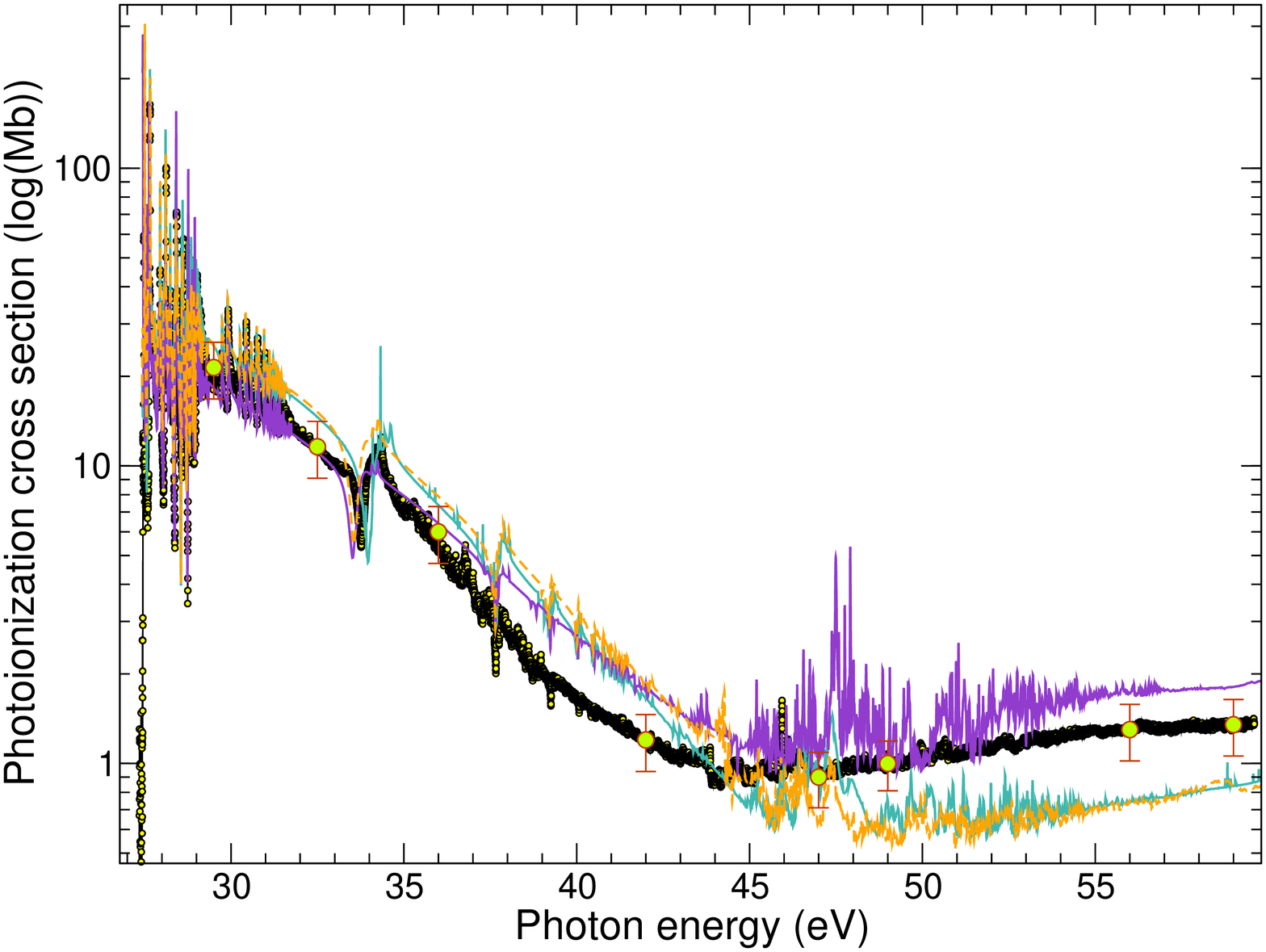}
\caption{Total photoionization cross section measured in Mb on a logarithmic scale as a function of photon energy in eV. All results display the initial ground state, statistically weighted $^2$P$^{\rm o}$, with $J=3/2$ and $J=1/2$ odd states to all allowed final states. A 10meV gaussian convolution at FWHM is applied to compare directly with experimental resolution for all theoretical calculations. The yellow circles, green circles with error bars, and solid turquoise line are the experimental results, absolute measurements at resonance free regions and theoretical calculations respectively, performed by \citet{2011PhRvA..84a3413C}. The dashed orange and solid purple lines represent our PBP and DARC3 calculations respectively. \label{fig:valence}}
\end{figure*}


\section{The photoionization calculations}\label{sec:rmatrix}
The photoionization process can be solved through the well established \textbf{R}-matrix theory \citep{1975JPhB....8.2620B} and we direct the reader towards \citet{2011rmta.book.....B} for a complete description. The photoionization cross section $\sigma$, defined in Megabarns (1 Mb = $10^{-18}$ cm$^{2}$) is then calculated from, 
\begin{equation}\label{eq:photo}
\sigma = \frac{4\pi a_0^2\alpha\omega}{3g_i} \sum (\Psi_f || \mathbf{D} || \Psi_i )
\end{equation}
and can be cast in both length and velocity gauges through the dipole moment operator $\mathbf{D}$, for the velocity gauge $\omega\rightarrow\omega^{-1}$. a$_0$ is the bohr radius, $\alpha$ the fine structure constant, $g_i$ the statistical weighting of the initial state and $\Psi_i$/$\Psi_f$ are initial/final state scattering wavefunctions. These transitions occur between the lowest initial odd states 3s$^2$3p$^5$ $^2$P$^{\rm{o}}$, with $J = 3/2$ and $J = 1/2$ to allowed even final states subject to the dipole selection rules. It has also been possible to produce the photoionization cross section originating from the initial excited state 3s3p$^6$ $^2$S$_{1/2}$ through the calculated dipole matrices. The total scattering wavefunctions are chosen such that they are accessible through appropriate coupling of the scattered electron with the target wavefunctions.

The \textbf{R}-matrix box was set at 13.28 a. u., which was sufficient to encompass our most diffuse orbital, the 4p. We also require 22 continuum orbitals to describe the outgoing electron at energies just above the L-shell. To ensure all resonant features have been properly resolved, a fine mesh of 60,000 points across the lower energy range 27-60eV has been adopted, with similar increments for the L-shell energy region. Both length and velocity forms of the dipole moment operator in Equation \ref{eq:photo} are in suitable agreement when describing the photoionization cross section. Up to 45eV the difference in results is within 20$\%$ and therefore all results are provided in the length gauge.


\section{Results}\label{sec:results}
Before embarking on the large scale DARC3 calculation we thought it prudent to investigate first the important properties and characteristics found in the photoionization cross section of Ar {\sc ii} in its ground state. In Figure \ref{fig:comparison}
we present the total photoionization cross section in Mb on a logarithmic scale as a function of photon energy in eV, from the initial ground Ar {\sc ii} $^2$P$^{\rm{o}}_{3/2}$ state to all allowed final states. Three calculations are presented in this figure; both the 209 level DARC1 and its contributions from the 3s$^2$3p$^4$ levels indexed as 1-5 in Table \ref{tab:energy} and the extended 257 level DARC2 calculation. Clearly Figure \ref{fig:comparison}
shows the importance of including at least the first five 3s$^2$3p$^4$ levels of Ar {\sc iii} in this photoionization calculation. The contributions from these levels dominates the total cross section up to a photon energy of approximately 50eV and all three calculations exhibit excellent agreement up to this point. It is essential, therefore, that an accurate description is achieved for the wavefunction representation of those low-lying levels. Above 50eV the additional levels associated with the more complex DARC1 and DARC2 models come into play and the cross section rises as we move to higher photon energies as more channels become accessible. Interestingly the inclusion of the additional 3s$^2$3p$^3$4d and 3s$^2$3p$^3$5s levels in the DARC2 model has little or no effect on the photoionization cross section produced by the DARC1 model up to 60eV, both datasets showing near perfect agreement. Therefore we do not retain these additional 3s$^2$3p$^3$4d and 3s$^2$3p$^3$5s configurations in our largest DARC3 calculation as can be seen from Table \ref{tab:calculations}.

\begin{figure*}
\includegraphics[scale=0.6, angle=-90]{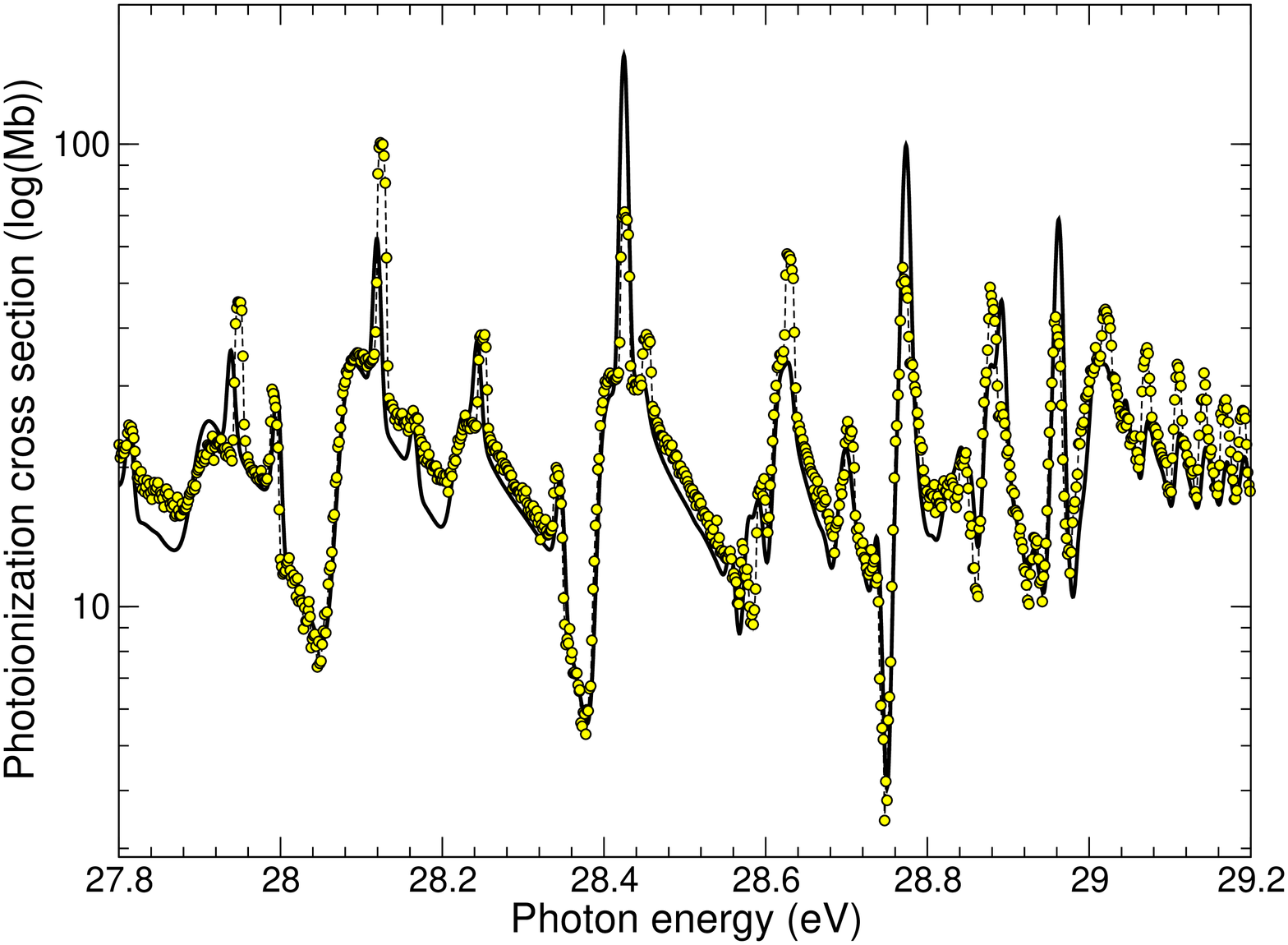}
\caption{Photoionization cross section measured in Mb on a logarithmic scale as a function of the photon energy between 27.8-29.2 eV just above threshold. The solid black line is the current statistically weighted, initial ground state, DARC3 calculation against the experimental values from \citet{2011PhRvA..84a3413C} represented by the yellow circles taken from Figure \ref{fig:valence}. \label{fig:valence_zoom}}
\end{figure*}

\begin{figure*}
\includegraphics[scale=0.58, angle=-90]{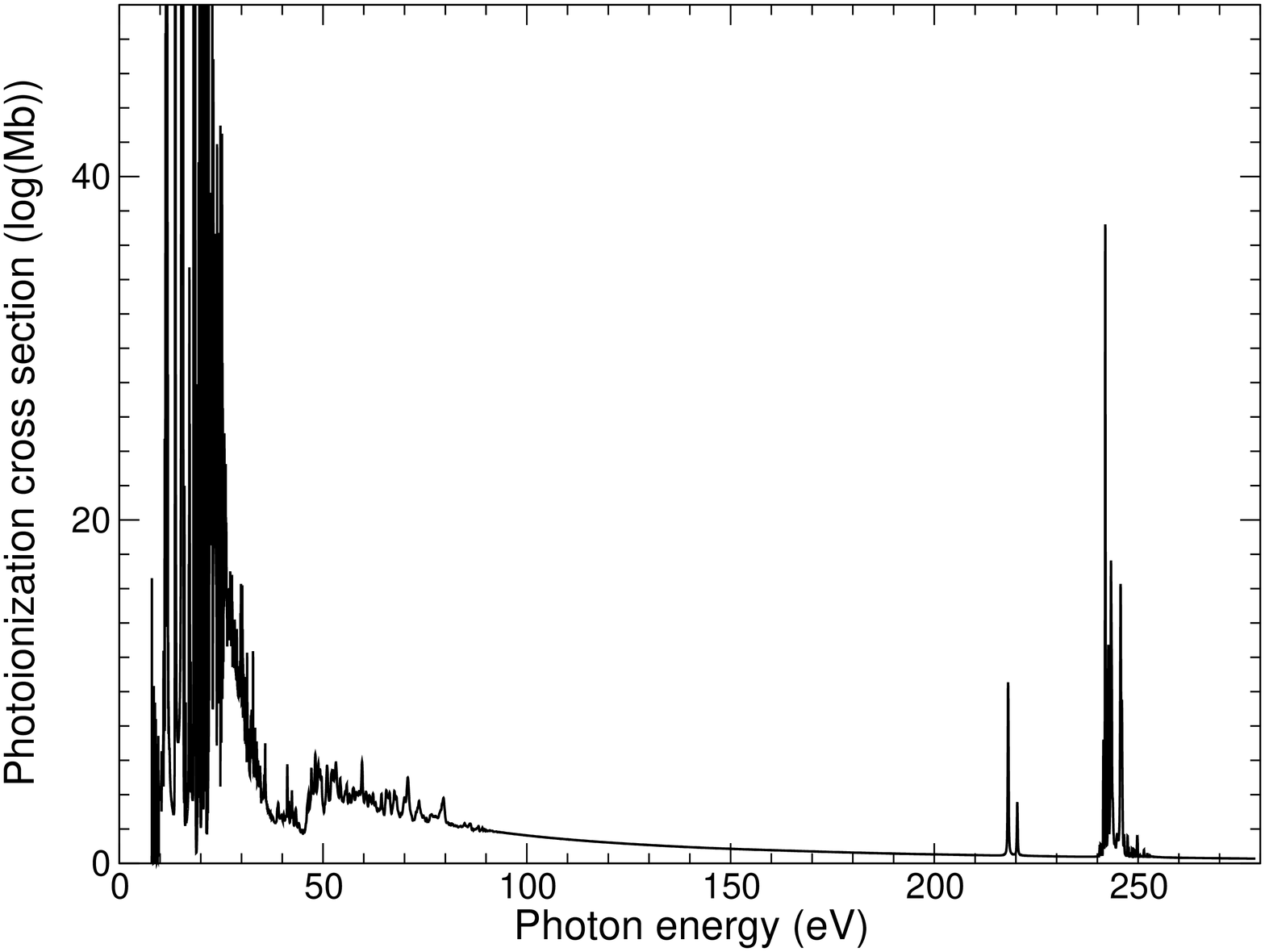}
\caption{Total ground state photoionization cross section measured in Mb as a function of the photon energy between 0-280eV. The transition is from the initial state 3s3p$^6$ $^2$S$_{1/2}$ to all allowed final states from the DARC3 model. \label{fig:excited}}
\end{figure*}

\subsection{Valence shell photoionization}
The only available data currently in the literature for valence shell photoionization of Ar {\sc ii} up to photon energies of 60eV is performed by \citet{2011PhRvA..84a3413C}. In this paper both theoretical and experimental cross sections are presented. Absolute cross sections are obtained from the merged beam technique at the Advanced Light Source (ALS) with a spectral resolution of 10meV. It was found that the primary ion beam contained a mixture of both $^2$P$_{3/2}^{\rm o}$ and $^2$P$_{1/2}^{\rm o}$ initial states. Hence the total cross section was presented as a statistical weighting of the odd parity $J=3/2$ ground and $J=1/2$ metastable states respectively. The accompanying theoretical cross sections presented by \citet{2011PhRvA..84a3413C} were evaluated using the Breit-Pauli \textbf{R}-matrix approach. A total of 48 $LSJ\pi$ fine-structure levels were included in the wavefunction representation with configurations 3s$^2$3p$^4$, 3s3p$^5$, 3p$^6$ and 3s$^2$3p$^2$3d$^2$. Some important correlation effects are thus omitted from this model such as levels associated with the 3s$^2$3p$^3$3d configuration and those arising from the lower $n=4$ complex. 

\begin{figure}
\includegraphics[scale=0.29, angle=-90]{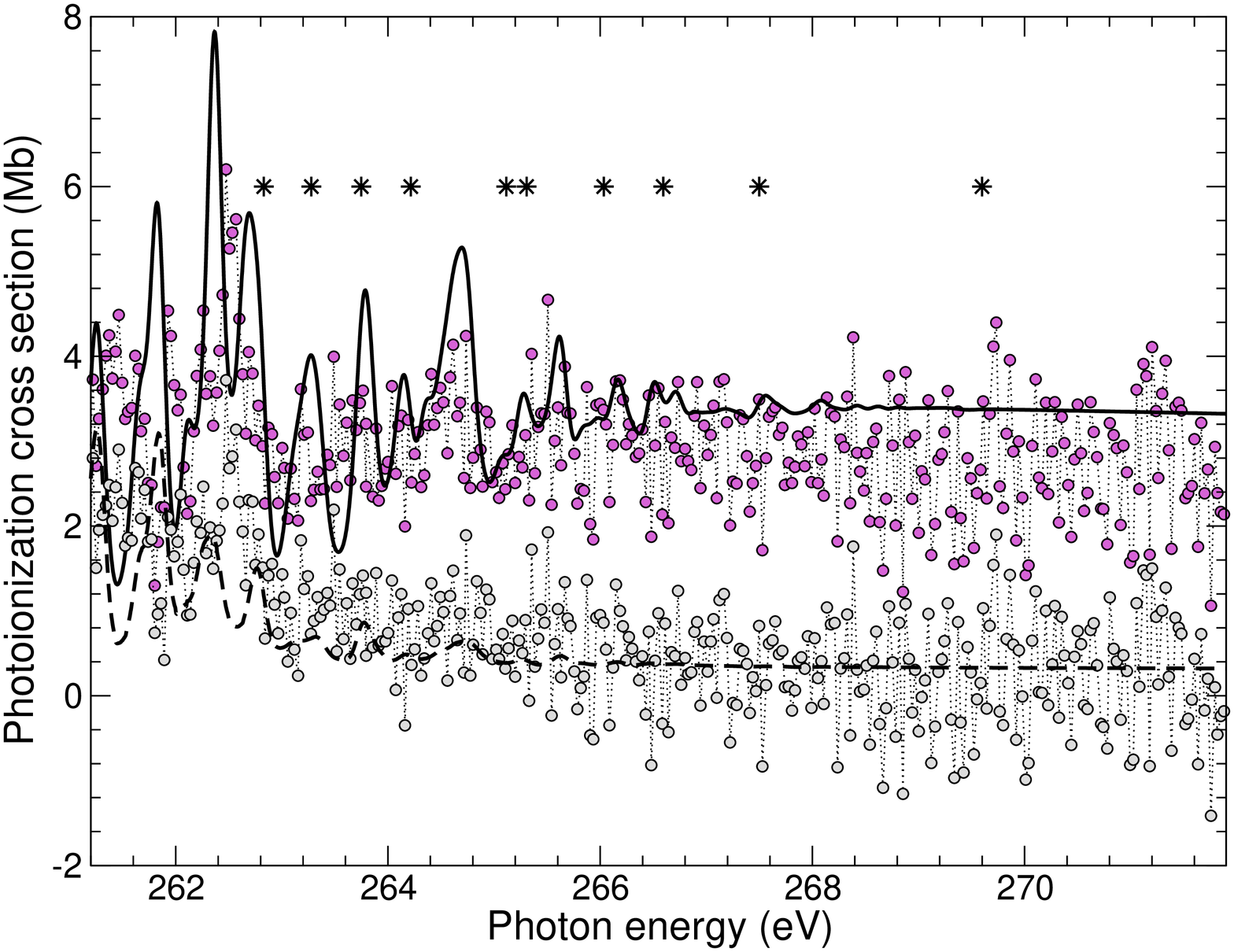}
\caption{The photoionization cross section is presented on a linear scale against photon energy in eV above 261.2eV. The solid black line represents our current DARC3 model convoluted at 140meV FWHM and the dashed black line is the contribution to the cross section from valence shell photoionization of the 3s and 3p. The grey circles are experimental values of \citet{2012PhRvA..85d3408B} for the single ionization channel and pink circles represent the total contribution. Each 2p$^5$3s$^2$3p$^5$ threshold is represented by an asterisk.\label{fig:s_and_d}}
\end{figure}

In order to compare with this data we present in Figure \ref{fig:valence} the total photoionization cross section from the initial $^2$P$^{\rm{o}}$ ground state of Ar {\sc ii} statistically weighted to the $J=3/2$ and $J=1/2$ states. We present two of our calculations in the figure, the most sophisticated DARC3 model and, in order to perform a direct comparison with the Breit-Pauli theoretical results of \citet{2011PhRvA..84a3413C}, the PBP 124 level model outlined in Table \ref{tab:calculations}. To match experimental resolving power, we convolute our total results with a 10meV gaussian profile at full-width half-maximum (FWHM). In addition, to replicate the target thresholds, we have shifted our threshold values recorded in Table \ref{tab:energy} to the experimental NIST values where possible, during the diagonalization of the Hamiltonian matrix. The remaining levels not contained in NIST are shifted by an average proportion to each corresponding angular and spin momentum state, which has little effect on the background and is meant only for consistency. This ensures that resonance features are properly positioned with respect to the observed thresholds, making a direct comparison with experiment more meaningful.

\begin{figure*}
\includegraphics[scale=0.6, angle=-90]{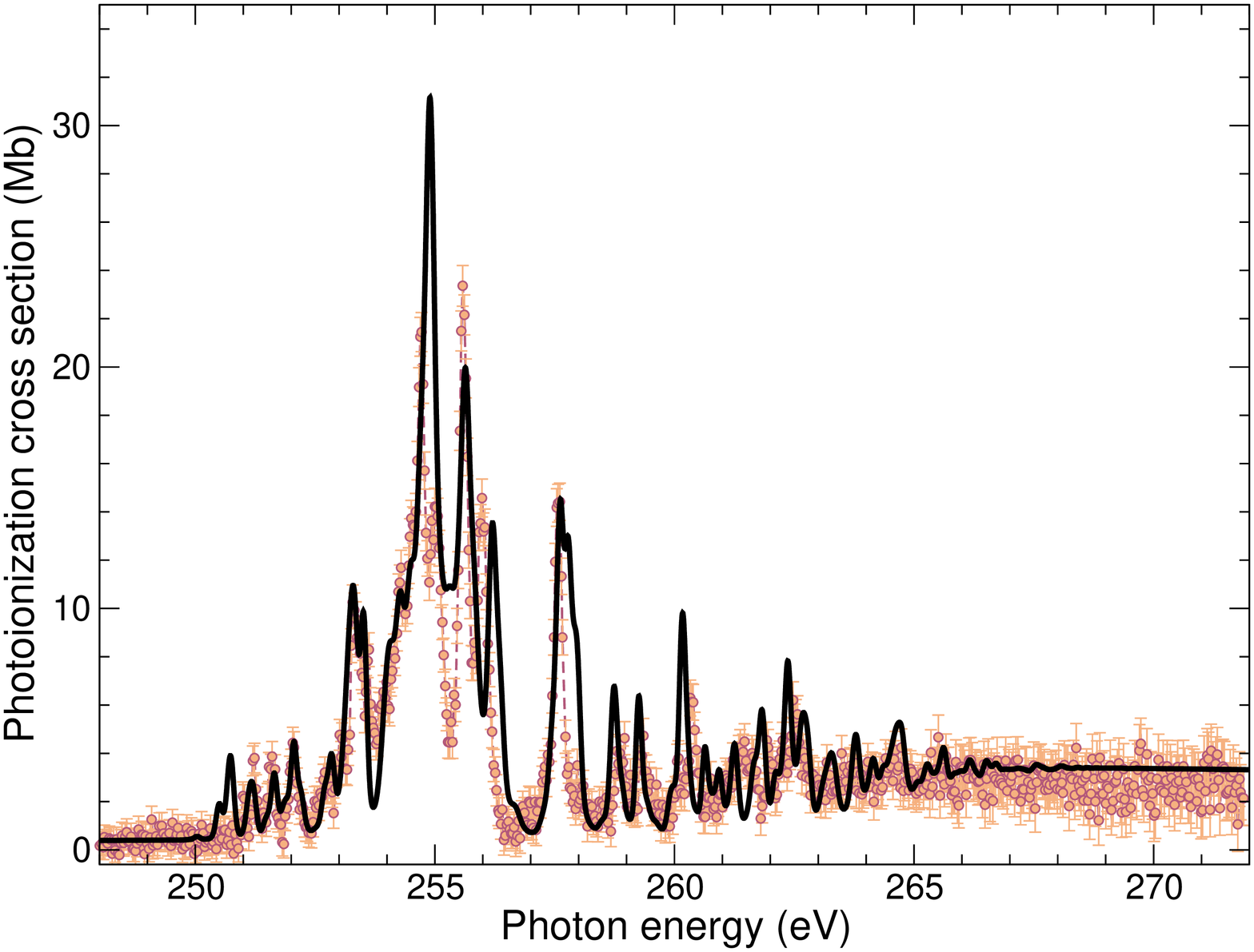}
\caption{Photoionization cross section measured in Mb on a logarithmic scale as a function of the photon energy in eV between 250-270eV. The circles are the experimental results from \citet{2012PhRvA..85d3408B} with error bars included. The solid black line represents our current, DARC3 model results for the statistically weighted initial ground state, convoluted at 140meV at FWHM. \label{fig:bigL}}
\end{figure*}

We can clearly see in Figure \ref{fig:valence} that the low energy region just above threshold is completely dominated by 3s$^2$3p$^5$ $\rightarrow$ 3s$^2$3p$^4nl$ transitions occurring at discrete energies prior to the ejection of an electron. This densely populated region of Rydberg resonances up to approximately 30eV is followed by a steep decline in the photoionization cross section forming the expected Cooper minimum around 45-50eV. This minimum is well known to appear in the spectra of noble gases \citep{1962PhRv..128..681C}. Above this minimum the cross section rises due to excitations from 3p $\rightarrow$ 3d transitions, before monotonically decreasing towards zero with increasing photon energy. 

Excellent agreement is evident between the 124 level PBP and the 48 level Breit-Pauli calculation of \citet{2011PhRvA..84a3413C}, for all photon energies up to 60eV. Note that the cross section in Figure \ref{fig:valence} is plotted on a log scale. Evidently the larger basis expansion of the present PBP evaluation, which includes the 4s, 4p and 4d orbitals, has minimal effect on the resulting photoionization cross section. Both of these Breit-Pauli evaluations, however, underestimate the cross section above roughly 45eV and lie considerably lower than the experimental measurements from ALS. The larger DARC3 evaluation, incorporating 557 fine-structure levels, gives much better agreement with experiment at photon energies above the Cooper minimum. This is partly due to the more substantial calculation, and also a more accurate description of the wavefunctions included. Both techniques in fact are known to reproduce similar results as shown in a study by \citet{2005JPhB...38.1667B}, showing that the average difference in effective collision strengths for Fe$^{14+}$ to be 6$\%$ between all transitions considered. The additional levels included, and the Rydberg resonances converging onto their thresholds, have the effect of raising the cross section above 45eV.

In order to further emphasize the excellent agreement between the DARC3 and the experimental measurements, we zoom in on the photon energy region just above threshold, from 27.8-29.2eV, in Figure \ref{fig:valence_zoom}. It is clearly evident that the disparities found between theory and experiment in this very narrow energy range are negligible and excellent conformity is achieved. This high level of agreement supports the accuracy of the DARC3 evaluation and we believe that these valence shell photoionization cross sections for the ground state of Ar {\sc ii} accurately reproduce the experimental spectrum. 

In Figure \ref{fig:excited} we present the total photoionization cross section for the process defined in Equation \ref{eq:excited}, photoionization from the lowest excited initial 3s3p$^6\; ^2$S$_{1/2}$ bound state of Ar {\sc ii} to all possible allowed final states of Ar {\sc iii}. These evaluations were carried out using the DARC3 model and present for the first time, cross sections for photoionization from an excited Ar {\sc ii} state. There are no other theoretical or experimental data with which we can compare in this figure. The cross section is presented as a function of the photon energy in eV which ranges from just above the ionization threshold to beyond the opening of the L-shell thresholds. The photoionization cross section tends towards zero with increasing energy, and it is only due to the inclusion of the additional 10 hole states in the DARC3 model do we witness contributions to the cross section at photon energies between 200-250 eV. 

\subsection{L-shell photoionization}
Calculations and experiment have been carried out at the L-shell energy region between 250-280eV by \citet{2012PhRvA..85d3408B} at the SOLEIL facility in France as described in Section \ref{sec:introduction}. All the results herein have been convoluted with a 140meV Gaussian profile at FWHM to match the spectral resolution of experiment. Similar to the valence shell comparisons, the initial ground state cross section is formed from a statistically weighted average of the contributions from the odd $J=3/2$ and $J=1/2$ partial waves. Due to time of flight between the ion source and interacting region, excited levels can populate the main ion beam. This leads to a possible inclusion of the initial 3s3p$^6$ $^2$S$_{1/2}$ bound state which may also contribute to the total cross section. In Figure \ref{fig:excited} we have already shown the immediate result of the lowest excited initial bound state transitions arising from the configuration 3s3p$^6$.

\begin{figure}
\includegraphics[scale=0.30, angle=-90]{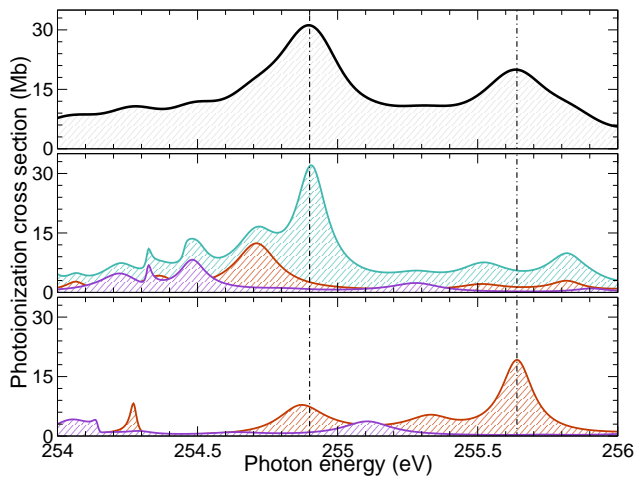}
\caption{The total convoluted FWHM at 140meV photoionization cross section between 254-256eV taken from Figure \ref{fig:bigL} highlighting the intense resonant peaks. The spectra is broken into the contributions from each dipole allowed symmetry from both initial (middle third) and metastable (bottom third) initial states according to their statistical weighting. These even $J=5/2, 3/2, 1/2$ partial symmetries are the solid turquoise, red and purple lines. The total (top third) summed contribution is presented by the solid black curve and the two dominate resonances are marked by the dashed line. \label{fig:res_peaks}}
\end{figure}

In order to compare with experiment, we have presented our results against various ionization channels from \citet{2012PhRvA..85d3408B} in Figure \ref{fig:s_and_d}. The timescale for Auger decay is much shorter than the time of flight required by the Argon ions after interaction with a photon, and therefore, the single ionization channel from experiment depicts the characteristics of photoionizing a valence electron. We can directly compare with this process in Figure \ref{fig:s_and_d} by omitting the contribution from the additional 10 target states annotated by asterisks. Both above and during these thresholds we expect a rise in the photoionization cross section as more channels are opened and become accessible. The total result obtained by DARC3 can be compared directly to the combination of both single and double ionization modes of experiment. We have neglected the error bars for both modes in order to visualise the results more clearly.

We now present in Figure \ref{fig:bigL} the photoionization cross section, on a linear scale, as a function of incident photon energy in eV across the L-shell threshold range from 250-270eV. Comparisons are made between the present DARC3 cross section and the measurements performed by \citet{2012PhRvA..85d3408B}. Clearly excellent agreement is evident between theory and experiment across the range considered, as the features and energy positions of the resonance profiles exhibit good agreement. We note that as we have employed orbitals optimized on the valence state photoionization, an energy shift of 7.5eV was required to match the experimental spectra to our current results. The theory clearly predicts this process to a high standard of accuracy and allows us to benchmark the quality of results obtained from experiment.

In an attempt to investigate the features further, we have broken down the spectrum in Figure \ref{fig:res_peaks} from the total into each of the allowed, final, even $J$ states $J=1/2$, $J=3/2$ and $J=5/2$. Clearly visible is the intense spike at $\approx 254.9$ eV which is dominated by transitions of the form, 2p $\rightarrow$ nd, ns which are engulfed by the convolution. The second strong peak at $\approx 255.65$ eV is visible mostly through the metastable initial state transition from another strong 2p $\rightarrow$ nd, J = 3/2 resonance. In reference to Figure \ref{fig:excited}, the cross section has already reached close towards zero in the photon energy range of interest and therefore any contribution to the total cross section from these initial excited bound states would result in a reduction to the intensity of each resonant state. 

This method of deconstructing the cross section is also important to identify which initial state has been photoionized during the experiment. It is clear however that the strongest profiles are not well isolated and therefore eliminates the possibility to further conduct any analysis on the weighted contributions. We therefore retain the statistical averaging of the ground state as our best result.

All resonances in this paper were identified using the technique detailed by \citet{1996JPhB...29.4529Q} and \citet{1998CoPhC.114..225Q}, which involves an analytic approach complementary to the \textbf{R}-matrix method. By exploiting multichannel quantum defect theory \citep{1983RPPh...46..167S}, each resonant state part of a Rydberg series has constant defect, $\mu$ for each effective $n$ quantum number defined by,
\begin{equation}\label{eq:resonance}
E_r = E_{n \rightarrow \infty} - \Big[\frac{Z-N}{n-\mu}\Big]^2
\end{equation}
where $E_r$ is the resonance energy converging to the target thresholds, $E_{n \rightarrow \infty}$. The overlapping nature of the resonant states makes it difficult to accurately evaluate resonance widths and assign each transition taking place. It is possible to deduce that the hole resonant states arise from 2p $\rightarrow nd, (n+1)s$ transitions for $n\ge3$, and correspond to the strongest peaks evident in Figure \ref{fig:res_peaks}.



\section{Conclusions}\label{sec:conclusions}
Photoionization cross sections have been produced for the three lowest states of Ar {\sc ii} from two independent \textbf{R}-matrix methods, {\sc darc} and {\sc bp}. Both evaluations differ considerably in size and sophistication. Energy levels obtained from the target wavefunctions have been shown to be in excellent agreement with previous theoretical works and also with the observed values contained in the NIST database. All photoionization spectra reported in this paper have been properly resolved with a very fine mesh of incident photon energies and comparisons have been made where possible. Excellent agreement is clearly evident between theory and experiment up to the L-shell energy region for all photon energies considered. Agreement has been obtained with the present DARC3 model which was shown to better reproduce experimental measurements for a number of energy ranges, particularly just above threshold and the Cooper minimum in the low energy section of the spectrum. The results pertaining to the energy region of 250-270eV are the first \textbf{R}-matrix calculations performed to date and clearly resolve the majority of resonance structure. This model thus represents the largest and most sophisticated evaluation of Ar {\sc ii} photoionization from the ground and first excited state, providing the most complete cross sections to date.


\section*{Acknowledgments}

The work presented in this paper has been supported by STFC through the grant ST/K000802/1. The authors would like to gratefully acknowledge the advice received from both J. M. Bizau \citep{2012PhRvA..85d3408B} and B. McLaughlin \citep{2011PhRvA..84a3413C} and also for providing their datasets presented in this paper.

\bibliographystyle{mn2e}
\bibliography{mybib.bib}

\end{document}